\documentclass{elsarticle}

\usepackage{lipsum}
\usepackage{mathrsfs}
\usepackage{amsmath,amsfonts,amssymb,mathtools}
\usepackage{dsfont}
\usepackage{graphicx}
\usepackage{hyperref}
\usepackage{verbatim}
\usepackage{slashed}
\usepackage[all]{xy}
\usepackage{xcolor}
\usepackage{subfig}
\usepackage{tikz}
\usetikzlibrary{shapes.geometric,arrows}
\usepackage{relsize}
\usepackage{caption}
\usepackage{float}
\usetikzlibrary{positioning}
\usepackage{geometry}
\usepackage[T1]{fontenc}
\usepackage{lmodern}
\usepackage{stackengine} 

\usepackage[utf8]{inputenc}

\hypersetup{
	colorlinks,
	citecolor=violet,
	linkcolor=blue,
	urlcolor=blue}

\csname @addtoreset\endcsname{equation}{section}
\DeclareMathAlphabet\mathbfcal{OMS}{cmsy}{b}{n}

\newcommand{\beq}{\begin{equation}}
	\newcommand{\eeq}{\end{equation}}
\newcommand{\bea}{\begin{eqnarray}}
	\newcommand{\eea}{\end{eqnarray}}
\newcommand{\ba}{\begin{array}}
	\newcommand{\ea}{\end{array}}
\newcommand{\bit}{\begin{itemize}}
	\newcommand{\eit}{\end{itemize}}

\newcommand{\complesso}{{\ \hbox{{\rm I}\kern-.6em\hbox{\bf C}}}}
\newcommand{\reale}{{\hbox{{\rm I}\kern-.2em\hbox{\rm R}}}}
\newcommand{\uno}{ \,  \raisebox{+0.14em}{{\hbox{{\rm \scriptsize ]}} \raisebox{-0.2em}{\kern-.8em\hbox{1}}}} \, }  

\newcommand {\be}{\begin{equation}}
	\newcommand {\ee}{\end{equation}}
\newcommand{\bm}{\boldsymbol}

\begin{document}
	
	\begin{frontmatter}
		
		
		
		\title{Accelerating Kerr-Taub-NUT spacetime in the low energy limit of heterotic string theory}
		
		
		\author{Haryanto M. Siahaan}
		\affiliation{organization={Jurusan Fisika,	Universitas Katolik Parahyangan},
			addressline={Jalan Ciumbuleuit 94}, 
			city={Bandung},
			postcode={40141}, 
			state={Jawa Barat},
			country={Indonesia}}
		
		\begin{abstract}
We construct a novel solution in the low energy limit of heterotic string theory describing accelerating charged and rotating black holes with NUT parameter. Some aspects of the new spacetime are discussed, such as locations of horizons, ergoregions, conic singularity, closed timelike curves, and area temperature product. We find some of the properties resemble the ones belong to accelerating Kerr-Newman-Taub-NUT spacetime. 
		\end{abstract}
		
		
		
		\begin{keyword}
			Accelerating black hole \sep Hassan-Sen \sep low energy heterotic string \sep area-temperature
			
			
			
		\end{keyword}

	\end{frontmatter}
	
	
	
	
\section{Introduction}
\label{sec:intro}

String theory is widely regarded as a coherent framework for explaining gravity on a quantum level. Consequently, discussions concerning the low-energy limits of string theory that encompass gravity remain a focal point of interest in the field of gravitational physics. In the context of heterotic string theory, the low-energy limit includes fundamental elements like the graviton, Abelian gauge field, dilaton, and second rank antisymmetric tensor fields. This low-energy limit gives rise to a specific black hole solution known as the Kerr-Sen black hole, originally introduced by Sen in his work  \cite{Sen:1992ua}. The Kerr-Sen black hole exhibits attributes such as mass, rotation, and electric charge, analogous to the well-established Kerr-Newman black hole within Einstein-Maxwell theory. Despite these similarities, distinctions between Kerr-Sen and Kerr-Newman black holes emerge in various aspects, including phenomena like strong light deflection in the Kerr-Sen background \cite{Gyulchev:2006zg} and the presence of hidden conformal symmetry \cite{Ghezelbash:2012qn}. These unique features, combined with the anticipation that string theory represents the ultimate explanation for all fundamental processes, motivate researchers to study deeper into the properties of Kerr-Sen black holes \cite{Siahaan:2015ljs, Siahaan:2015xna, Huang:2017whw, Uniyal:2017yll, Liu:2018vea}.

In the process of deriving the Kerr-Sen solution, Sen employed the Hassan-Sen transformation \cite{Hassan:1991mq}, which is rooted in the symmetries inherent in the low-energy limit of the heterotic string theory action. The Hassan-Sen transformation enables the mapping of a known solution in the theory, referred to as the seed solution, to another solution belonging to the same theory. However, it is customary to select a relatively simple seed solution, as this mapping process can lead to intricate expressions for all the incorporated fields. In the case of the Kerr-Sen solution \cite{Sen:1992ua}, Sen utilized the Kerr metric as the seed solution, resulting in a set of non-trivial solutions for all the fields in the theory. Subsequently, this solution-generation method was employed in \cite{Johnson:1994ek}, where the authors adopted the non-rotating Taub-NUT spacetime as their seed solution.

In the last decades, accelerating Taub-NUT spacetime has been discussed by several authors \cite{Astorino:2016xiy,Appels:2016uha,Appels:2017xoe,Astorino:2016ybm,Anabalon:2018ydc,Podolsky:2020xkf,Podolsky:2021zwr,Podolsky:2022xxd,Barrientos:2023tqb,Siahaan:2018qcw,Barrientos:2023dlf}. In ref. \cite{Podolsky:2020xkf}, the authors provide a convenient way in expressing the accelerating Taub-NUT metric and generalized to the case with rotation and electric charge in \cite{Podolsky:2021zwr}. Furthermore, accelerating Taub-NUT spacetimes with cosmological constant were presented in \cite{Podolsky:2022xxd}, and a convenient form of accelerating and rotating Taub-NUT spacetime metric is given in \cite{Barrientos:2023tqb}. This series of works reflect the importance of accelerating Taub-NUT solution for our understanding of Einstein theory. Note that accelerating Taub-NUT spacetime discussed in \cite{Podolsky:2020xkf} solves the vacuum Einstein equation, and the one equipped with electric charge belongs to Einstein-Maxwell system \cite{Podolsky:2021zwr,Barrientos:2023tqb}.

In the domain of low-energy heterotic string theory, we have come across solutions resembling the accelerating Kerr-Newman and Kerr-Newman-Taub-NUT spacetimes in previous works \cite{Siahaan:2018qcw, Siahaan:2019kbw}. Now, the advent of a more convenient form for the accelerating-Kerr-Newman-Taub-NUT solution, as presented in \cite{Podolsky:2021zwr, Barrientos:2023tqb}, has inspired us to seek a similar solution within the framework of low energy heterotic string theory. Our objective here is to not only acquire this analogous solution but also to explore its properties. To achieve this, we employ the Hassan-Sen transformation, starting with the accelerating Kerr-Taub-NUT spacetime as seed solution.

Organization of this paper is as follows. In the next section, we present brief reviews on the accelerating Kerr-Newman-Taub-NUT spacetime and the solution generating method in the low energy of heterotic string theory \cite{Sen:1992ua}. Section \ref{sec.AccKSTN} presents the accelerating Kerr-Sen-Taub-NUT solution, and its properties are discussed in the following section. Finally, we give some discussions. In this paper, we adopt natural units where $c$, $G$, and $\hbar$ are all set to unity.

\section{Brief reviews}\label{sec.review}

\subsection{Accelerating Kerr-Newman-Taub-NUT spacetime}

Accelerating Kerr-Newman-Taub-NUT (AKNTN) solution is an electrovacuum system that solves the Einstein-Maxwell equations
\be \label{eq.EinsteinMaxwell}
R_{\mu\nu} = 2 F_{\mu \beta} F_{\nu}^\beta - \frac{1}{2} g_{\mu\nu} F_{\alpha \beta} F^{\alpha \beta} \,
\ee 
and the source-free condition 
\be \label{eq.sourcefree}
\nabla_\mu F^{\mu \nu} = 0\,,
\ee 
where $R_{\mu\nu}$ is the Ricci tensor and $F_{\mu\nu} = \partial_\mu A_\nu - \partial_\nu A_\mu$ is the field-strength tensor of electromagnetic field. The line element of AKNTN solution can be expressed as \cite{Podolsky:2021zwr}
\be \label{metric.AKNTN}
ds^2  = \frac{1}{{\Omega ^2 }}\left( { - \frac{{\Delta _r }}{\Sigma }\left[ {dt - Z_x d\phi } \right]^2  + \Sigma \left[ {\frac{{dr^2 }}{{\Delta _r }} + \frac{{dx^2 }}{{P\Delta _x }}} \right] + \frac{{P\Delta _x }}{\Sigma }\left[ {adt - Z_r d\phi } \right]^2 } \right)\,,
\ee 
where
\[
\Omega  = 1 - \frac{{bar}}{{a^2  + l^2 }}\left( {l + ax} \right)\,,
\]
\[
\Sigma  = r^2  + \left( {l + ax} \right)^2 \,,
\]
\[
P = \left( {1 - \frac{{bar_ +  }}{{a^2  + l^2 }}\left( {l + ax} \right)} \right)\left( {1 - \frac{{bar_ -  }}{{a^2  + l^2 }}\left( {l + ax} \right)} \right)\,,
\]
\[
\Delta _r  = \left( {r - r_ +  } \right)\left( {r - r_ -  } \right)\left( {1 + \frac{{ba\left( {a - l} \right)r}}{{a^2  + l^2 }}} \right)\left( {1 - \frac{{ba\left( {a - l} \right)r}}{{a^2  + l^2 }}} \right)\,,
\]
\[
Z_r = r^2 + (a+l)^2 \,,
\]
\[
Z_x = a \Delta_x + 2l (1-x)\,,
\]
and $\Delta_x = 1-x^2$. Here, we have used the coordinate $x^\mu = \left[t,r,x=\cos\theta,\phi\right]$. In equations above, $r_+$ and $r_-$ denote the outer and inner black hole horizon radii, respectively, which are given by
\be 
r_\pm = M \pm \sqrt{M^2+l^2 -a^2 - Q^2}\,.
\ee

It is commonly recognized that the parameters $M$, $a$, $Q$, and $l$ represent the mass, rotation, electric charge, and NUT charge characteristics of an object described by the AKNTN solution. In addition to these parameters, there is an acceleration parameter denoted as $b$ which signifies the acceleration between two black holes responsible for generating the AKNTN spacetime. Specifically, we use the notation $r_\pm$ to represent the black hole horizons. Beyond these horizons, there exists another radius that results in the metric (\ref{metric.AKNTN}) becoming singular. This can be observed when $\Delta_r$ becomes zero at a certain point
\[
r_b = \pm \frac{a^2+l^2}{ba \left(a \pm l\right)}\,.
\]
This horizon is regarded as the acceleration horizon. The potential vector accompanying the metric above in solving eqs. (\ref{eq.EinsteinMaxwell}) and (\ref{eq.sourcefree}) can be expressed as
\[
A_\mu  dx^\mu   = \frac{{Qr}}{\Sigma }\left( {dt - Z_x d\phi } \right)\,.
\]

\subsection{Hassan-Sen transformation in the low energy limit of heterotic string theory}\label{revHS}

In \cite{Sen:1992ua}, Sen utilized a transformation to convert the Kerr solution into a spacetime configuration representing a charged and rotating black hole. This transformation was applied within the constraints of the low-energy limit of heterotic string theory. The technique used for this transformation is known as the Hassan-Sen transformation, originally introduced by Hassan and Sen in their work \cite{Hassan:1991mq}. The Hassan-Sen transformation is a mathematical method that establishes a correspondence between two sets of fields. It maps the first set, denoted as $\left\{ {g_{\mu \nu }, A_\mu  ,\Phi, B_{\mu \nu } } \right\}$, to a second set, denoted as $\left\{ {{  g}_{\mu \nu }' ,{  A}_\mu ' ,{  \Phi} ',{  B}_{\mu \nu }' } \right\}$, ensuring that both sets satisfy the equations of motion derived from the action
\be\label{action.het}
S = \int {d^4 x} \sqrt { - g} e^{ - \Phi } \left( {R + \left( {\nabla \Phi } \right)^2  - \frac{1}{8}F_{\mu \nu } F^{\mu \nu }  - \frac{1}{{12}}H_{\alpha \beta \gamma } H^{\alpha \beta \gamma } } \right)\,.
\ee
Above, the third rank tensor is defined as \be 
H_{\alpha \beta \gamma }  = \partial _\alpha  B_{\beta \gamma }  + \partial _\gamma  B_{\alpha \beta }  + \partial _\beta  B_{\gamma \alpha }  - \frac{1}{4}\left( {A_\alpha  F_{\beta \gamma }  + A_\gamma  F_{\alpha \beta }  + A_\beta  F_{\gamma \alpha } } \right)
\ee 
where the field-strength tensor $F_{\beta \gamma }$ adopts precisely the same structure as it does in the Einstein-Maxwell theory. The transformation relies on a constant parameter, which, in the transformed field solution, contributes to the electric charge.

If one considers a stationary and axially symmetric vacuum solution as the initial state for the Hassan-Sen transformation, the procedure can be further explained as follows. Assume the seed field is denoted as $\left\{{\tilde g}_{\mu\nu}\right\}$ and comprises the components
\be\label{metric0}
ds^2  = {\tilde g}_{tt} dt^2  + 2{\tilde g}_{t\phi } dtd\phi  + {\tilde g}_{\phi \phi } d\phi ^2  + {\tilde g}_{rr} dr^2  + {\tilde g}_{xx} dx^2 \,.
\ee 
Above, the metric functions are $r$ and $x$ dependent. The action for vacuum system is given by
\be\label{action.vacuum.Einstein}
S = \int {d^4 x\sqrt { - {\tilde g}} {\tilde R}} \,.
\ee 
In the last equation, $\tilde R$ is the Ricci scalar for the spacetime with tensor metric ${\tilde g}_{\mu\nu}$. Intuitively, $\tilde g$ is the determinant of ${\tilde g}_{\mu\nu}$. Now the Hassan-Sen transformation brings the seed field $\left\{{\tilde g}_{\mu\nu}\right\}$ to  $\left\{ {{ g}_{\mu \nu } ,A_\mu  ,\Phi ,B_{\mu \nu } } \right\}$ obeying the equations of motion obtained from the low energy heterotic string theory action in eq. (\ref{action.het}), namely
\be \label{eqG}
{ R}_{\mu \nu } - \frac{1}{2} { g}_{\mu\nu} \left[{ R} - \left( {\left( {\nabla \Phi } \right)^2  - 2\nabla ^2 \Phi } \right)\right] + \nabla _\mu  \nabla _\nu  \Phi    = \frac{1}{4}\left[ {F_{\mu \alpha } F_\nu ^\alpha   + H_{\mu \alpha \beta } H_\nu ^{\alpha \beta }  - \frac{1}{2}g_{\mu \nu } \left( {\frac{{F^2 }}{2} + \frac{{H^2 }}{3}} \right)} \right]\,,
\ee 
\be \label{eqPh}
\left( {\nabla \Phi } \right)^2  - 2\nabla ^2 \Phi  = R - \frac{{F^2 }}{8} - \frac{{H^2 }}{{12}}\,,
\ee
\be \label{eqA}
\nabla _\mu  F^{\mu \nu }  = F^{\mu \nu } \partial _\mu  \Phi  + \frac{1}{2}F_{\alpha \beta } H^{\nu \alpha \beta } \,,
\ee
and
\be \label{eqB}
\nabla _\alpha  H^{\alpha \mu \nu }  = H^{\alpha \mu \nu } \partial _\alpha  \Phi \,.
\ee

To see how explicitly Hassan-Sen transformation works, let us first introduce a second rank tensor second-rank tensor formed from the vector potential, metric tensor, and the $B_{\mu\nu}$ field as follows
\be
K_{\mu \nu }  =  - g_{\mu \nu }  - \frac{1}{4}A_\mu  A_\nu   - B_{\mu \nu }\,.
\ee 
The elements of $K_{\mu\nu}$ form the matrix $\bf{K}$. Next, we can represent the flat spacetime metric as
\be 
{\eta}_{\mu\nu} = {\rm diag}\left(1,1,1,-1\right)
\ee 
the components of which can be represented in the form of the matrix $\bm \eta$. Moreover, we can create a $9\times 9$ matrix as follows
\[
{\bf M} = \left( {\begin{array}{*{20}c}
	{\left( {{\bf K}^T  - {\bm \eta} } \right){\bf g}^{-1} \left( {{\bf K} - {\bm \eta} } \right)} & {\left( {{\bf K}^T  - {\bm \eta} } \right){\bf g}^{-1} \left( {{\bf K} + {\bm \eta} } \right)} & { - \left( {{\bf K}^T  - {\bm \eta} } \right){\bf g}^{-1} {\bf A}}  \\
	{\left( {{\bf K}^T  + {\bm \eta} } \right){\bf g}^{-1} \left( {{\bf K} - {\bm \eta} } \right)} & {\left( {{\bf K}^T  + {\bm \eta} } \right){\bf g}^{-1} \left( {{\bf K} + {\bm \eta} } \right)} & { - \left( {{\bf K}^T  + {\bm \eta} } \right){\bf g}^{-1} {\bf A}}  \\
	{ - {\bf A}^T {\bf g}^{-1} \left( {{\bf K} - {\bm \eta} } \right)} & { - {\bf A}^T {\bf g}^{-1} \left( {{\bf K} + {\bm \eta} } \right)} & {{\bf A}^T {\bf g}^{-1} {\bf A}}  \\
	\end{array}} \right)
\]
where the matrix $\bf A$ contains the component of potential vector $A_\mu$ and $\bf g$ is the matrix form of metric tensor $g_{\mu\nu}$. Interestingly, the action (\ref{action.het}) is invariant under the transformation \cite{Hassan:1991mq}
\be \label{eq.HassaSenMatrix}
{\bf { M}} \to {\bf M}' = {\bf \Theta { M}\Theta} ^T \,,
\ee
where
\be 
{\bf \Theta}  = \left( {\begin{array}{*{20}c}
		{{\rm I}_{7 \times 7} } & {...} & {...}  \\
		{...} & {\cosh \alpha  } & {\sinh \alpha }  \\
		{...} & {\sinh \alpha } & {\cosh \alpha  }  \\
\end{array}} \right)\,,
\ee
provided that the dilation field is mapped as
\be 
{ \Phi}  \to \Phi '  = \Phi  + \frac{1}{2}\ln \frac{{\det {\bf g}'}}{{\det {\bf { g}}}}\,.
\ee
In the last equation, ${\bf g}'$ is understood as the one that constructs ${\bf M}'$. Note that the Hassan-Sen transformation depends on the constant $\alpha$. Setting $\alpha \to 0$ yields the transformation matrix $\bf \Theta$ to become a nine dimensional identity matrix ${\rm I}_{9 \times 9}$, which is related to identity transformation in eq. (\ref{eq.HassaSenMatrix}). 

Now let us see how the seed metric ${\tilde g}_{\mu\nu}$ is related to the set of fields obeying the equations of motion in the low energy limit of heterotic string theory. It can be found that the tensor metric components in the string frame is given by \cite{Siahaan:2018qcw}
\[
g_{tt} = \frac{{\tilde g}_{tt}}{\Lambda^2}~~,~~g_{rr} = {\tilde g}_{rr}~~,~~g_{xx} = {\tilde g}_{xx}~~,~~g_{t\phi} =\frac{\cosh^2\alpha}{\Lambda^2} {\tilde g}_{t\phi}\,,
\]
and
\be \label{metric.HassanSen}
g_{\phi\phi} = \frac{\left({\tilde g}_{\phi\phi} {\tilde g}_{tt} - {\tilde g}_{t\phi}^2\right)\Lambda^2 + {\tilde g}_{t\phi}^2 \cosh^2 \alpha}{\Lambda^2 {\tilde g}_{tt}}\,.
\ee
On the other hand, the gauge potential and dilaton field are given by 
\be \label{vector.HassanSen}
A_\mu  dx^\mu   = \frac{{\sinh (2\alpha) }}{\Lambda }\left( {\left( {1 + {\tilde g}_{tt} } \right)dt + {\tilde g}_{t\phi } d\phi } \right)\,,
\ee
\be \label{dilaton.HassanSen}
\Phi  =  - \ln \Lambda \,,
\ee 
respectively. Finally, the only non-vanishing components of the second rank tensor field is
\be \label{secondrank.HassanSen}
B_{t\phi }  =  - B_{\phi t}  = \frac{{\tilde g}_{t\phi } ~{\sinh^2 \alpha  }}{\Lambda }\,.
\ee
In equations above, $\Lambda  = 1 + \sin ^2 \alpha \left( {1 + \tilde g_{tt} } \right)$.

\section{Accelerating Kerr-Sen-Taub-NUT spacetime}\label{sec.AccKSTN}

In \cite{Sen:1992ua}, Sen employed the Kerr metric as the starting point in the Hassan-Sen transformation process discussed earlier. This led to the well-known Kerr-Sen solution, a topic that has been extensively examined in scientific literature. If we use the accelerating Kerr solution, also referred to as the rotating C-metric, as the seed, we obtain the accelerating Kerr-Sen spacetime, as described in \cite{Siahaan:2018qcw}. On the other hand, employing the Kerr-Taub-NUT solution as the seed results in the Kerr-Sen-Taub-NUT spacetime, as discussed in \cite{Siahaan:2019kbw}.

In this section, we consider a more general vacuum solution as the initial state, specifically the accelerating Kerr-Taub-NUT (AKTN) spacetime. This corresponds to the limit where the electric charge $Q$ approaches zero in the metric (\ref{metric.AKNTN}). In this limit, the only modified function is $\Delta_r$, which can be expressed in a similar form as in (\ref{metric.AKNTN})
\be 
\Delta _r  = \left( {r - r_ +  } \right)\left( {r - r_ -  } \right)\left( {1 + \frac{{ba\left( {a - l} \right)r}}{{a^2  + l^2 }}} \right)\left( {1 - \frac{{ba\left( {a - l} \right)r}}{{a^2  + l^2 }}} \right) \,,
\ee
but with the horizons
\be 
r_\pm = m \pm \sqrt{m^2+l^2-a^2}\,.
\ee 
In equations above, $a$, $m$, $l$, and $b$ are the rotational, mass, NUT, and acceleration parameters, respectively. Indeed, as a vacuum solution, this metric solves the $R_{\mu\nu}=0$ equation, and the accompanying gauge potential vanishes. 

In the string frame, by using eq. (\ref{metric.HassanSen}), the spacetime metric components as the result of Hassan-Sen transformation to the AKTN solution can be written as
\[
{ g}_{tt} = -\frac{\left(\Delta_r - P a^2 \Delta_x\right)}{\Lambda^2 \Omega^2 \Sigma}\,,
\]
\[
{ g}_{t\phi} = -\frac{\cosh^2\alpha\left(\Delta_r Z_x - P a \Delta_x Z_r\right)}{\Lambda^2 \Omega^2 \Sigma} \,,
\]
\be \label{NewMetric}
{ g}_{\phi\phi} =
\frac{{\Delta _r \Delta _x P\left( {Z_r  - aZ_x } \right)^2 }}{{\Omega ^2 \Sigma \left( {\Delta _r  - a^2 P\Delta _x } \right)}} - \frac{\cosh^4\alpha~  \left(\Delta_r Z_x - P a \Delta_x Z_r\right)^2}{\left(\Delta_r - P a^2 \Delta_x\right)\Sigma \Omega^2 \Lambda^2}
\ee 
whereas ${ g}_{rr}$ and ${ g}_{xx}$ are the same with those of seed solution in eq. (\ref{metric.AKNTN}) with $Q=0$. The metric (\ref{NewMetric}) represents a new solution arising in the low-energy limit of heterotic string theory, and it will be referred to as the Accelerating Kerr-Sen-Taub-NUT (AKSTN) metric. It serves as a counterpart to the AKNTN solution in equation (\ref{metric.AKNTN}). On the other hand, the non-gravitational fields within the AKSTN solution can be expressed as follows
\be 
A_\mu  dx^\mu   = \frac{{\sinh \left( {2\alpha } \right)}}{\Lambda }\left[ {\frac{{\left( {\Omega ^2 \Sigma  - \Delta _r  + Pa^2 \Delta _x } \right)}}{{\Omega ^2 \Sigma }}dt + \frac{{\left( {\Delta _r Z_x  - Pa\Delta _x Z_r } \right)}}{{\Omega ^2 \Sigma }}d\phi } \right]\,,
\ee 
\be 
\Phi  = -\ln \Lambda\,,
\ee 
and
\be 
B_{t\phi }  =  - B_{\phi t}  = \frac{{\sinh ^2 \alpha \left( {\Delta _r Z_x  - Pa\Delta _x Z_r } \right)}}{{\Omega ^2 \Sigma \Lambda }}\,.
\ee
Note that explicitly we have
\be 
\Lambda = \frac{{\left( {\Omega ^2 \Sigma  - \Delta _r  + Pa^2 \Delta _x } \right)\sinh ^2 \alpha  + \Omega ^2 \Sigma }}{\Omega ^2 \Sigma}\,.
\ee 
The set of fields $\left\{g_{\mu\nu},A_{\mu},\Phi,B_{\mu\nu} \right\}$ above solve the set of equations (\ref{eqG}), (\ref{eqA}), (\ref{eqPh}), and (\ref{eqB}). Nevertheless, for the purpose of exploring the spacetime's properties in the upcoming section, we will utilize the metric's Einstein frame, defined as
\be 
ds_{ E}^2 = G_{\mu\nu} dx^\mu dx^\nu = e^{ -\Phi} { g}_{\mu\nu} dx^\mu dx^\nu\,.
\ee 
Explicitly, the relation between the component of metric tensor in Einstein and string frames is 
\be \label{metric.AKSTN.Einstein}
G_{\mu\nu} = \Lambda { g}_{\mu\nu} \,.
\ee 

Certainly, the investigations concerning conserved quantities within a spacetime characterized by NUT and acceleration parameters have been a lively research area over the past few years. Nonetheless, for our present discussion, we will focus on the mass associated with the Kerr-Sen spacetime, as presented in \cite{Sen:1992ua}. The mass, electric charge, and angular momentum are determined as follows
\be 
M = \frac{m \left(1+\cosh \alpha\right)}{2}~~,~~~Q = \frac{m \sinh \alpha}{\sqrt{2}}\,,
\ee 
and $J = Ma$, respectively. Using these quantities, the inner and outer horizons of black hole in the AKSTN spacetime can be written as
\be 
r_{\pm} = \left(M-\frac{Q^2}{2M}\right) \pm \sqrt{\left(M-\frac{Q^2}{2M}\right)^2+l^2-a^2}\,.
\ee 
Accordingly, the naked singularity in AKSTN spacetime can be achieved if the black hole condition
\be \label{eq.BHcondition}
a^2 \ge \left(M-\frac{Q^2}{2M}\right)^2 +l^2
\ee 
is violated. It is obvious that the inner and outer horizons coincide, i.e. at the extremal state, for $a^2 = (M-\frac{Q^2}{2M})^2 +l^2$.

Before we proceed into specific aspects of the AKSTN solution, it is worth mentioning some subclasses of this spacetime. A diagram illustrating the interrelationships among the various members within the AKSTN family is provided in Figure \ref{fig:map}.
The Kerr-Sen-Taub-NUT spacetime can be obtained by setting the acceleration parameter $b$ to zero. Discussions related to this spacetime are presented in \cite{Siahaan:2019kbw}. On the other hand, the accelerating Kerr-Sen spacetime is achieved when the NUT parameter $l$ is set to zero. This solution was derived in \cite{Siahaan:2018qcw} by employing the Hassan-Sen transformation on the accelerating Kerr metric. Finally, 
In the limit $a\to 0$ within the AKSTN solution, we recover the non-accelerating charged Taub-NUT solution as discussed in the context of string theory by Johnson in \cite{Johnson:1994ek}. This reflects the known limitation that the accelerating Taub-NUT solution cannot be simply obtained by setting the rotational parameter $a$ to zero in the AKTN solution, as highlighted in \cite{Podolsky:2021zwr}. Note that the AKTN solution presented in \cite{Podolsky:2021zwr} serves as the seed for the Hassan-Sen transformation used in this work to derive the AKSTN solution. Consequently, the challenge of obtaining the non-rotating, accelerating charged Taub-NUT solution from its rotating counterpart persists in the seed solution and is carried over to the Hassan-Sen transformed solution.

\tikzstyle{rect} = [draw,rectangle, fill=white!20, text width =3cm, text centered, minimum height = 1.5cm,scale=0.9]
\begin{figure}
	\vspace{-1cm}
	
	\begin{center}
		
		\begin{tikzpicture}
		
		\hspace{-0.1cm}    \node[rect,scale=1.2,label=above :{},text width=4.5cm,node distance=1.5cm,line width=1.3pt](AKSTN){Accelerating Kerr-Sen-Taub-NUT\\ $\left\{M,a,l,Q,b\right\}$};
		\node[rect,scale=1.2,anchor=north,label=above :{~~~~~~~~~~~~~~~~~~~\cite{Siahaan:2018qcw}},
		text width=4.5cm,below of=AKSTN,node distance=3.3cm,line width=1.3pt](AKS){Accelerating Kerr-Sen\\ $\left\{M,a,Q,b\right\}$};
		\node[rect,scale=1.1,anchor=north,label=above :{\cite{Siahaan:2019kbw}},text width=4.2cm, left of= AKS, node distance=6cm](KSTN){Kerr-Sen-Taub-NUT\\ $\left\{M,a,l,Q\right\}$};
		\node[rect,scale=1.1,anchor=north,label=above:{\cite{Johnson:1994ek}},text width=3.5cm, right of= AKS, node distance=6cm](GMGHSNUT){Stringy charged Taub-NUT\\ $\left\{M,l,Q\right\}$};
		\node[rect,scale=1.2,anchor=north,label=above :{\cite{Sen:1992ua}~~~~~~~~~~~~~~~~},text width=2cm, below left of=AKS,node distance=6.7cm](Kerr-Sen){Kerr-Sen\\ $\left\{M,a,Q\right\}$};
		\node[rect,scale=1.2,anchor=north,label=above :{},text width=3.5cm, below of= AKS,node distance=4cm,line width=1.3pt](AGMGHS){Accelerating Gibbons-Maeda-Garfinkle-Horowitz-Strominger\\ $\left\{M,Q,b\right\}$};
		\node[rect,scale=1.2,anchor=north,label=above :{},text width=2cm, below of= GMGHSNUT,node distance=7.7cm](Taub-NUT){Taub-NUT\\ $\left\{M,l\right\}$};
		\node[rect,scale=1.2,anchor=north,below of=Kerr-Sen,node distance=3.2cm,label=above left:{},text width=2cm](Kerr){Kerr\\ $\left\{M,a\right\}$};
		\node[rect,scale=1.2,anchor=north,below of=AGMGHS,node distance=5.2cm,line width=1.3pt, label=above :{\cite{Gibbons:1987ps,Garfinkle:1990qj}~~~~~~~~~~~~~~~~~~~~~},text width=3.4cm](GMGHS){Gibbons-Maeda-Garfinkle-Horowitz-Strominger\\ $\left\{M,Q\right\}$};
		\node[rect,scale=1.2,anchor=north,below of=GMGHS,node distance=3.4cm, label=above :{ },text width=3cm](Sch){Schwarzschild\\ $\left\{M\right\}$};

		\draw[->] (AKSTN) -- node [right] {$l$ = 0}(AKS);
		\draw[->] (AKS) -- node [left,near start] {$b$ = 0} (Kerr-Sen);
		\draw[->] (AKSTN) -- node [right,near start] {}(GMGHSNUT);
		\draw[->] (GMGHSNUT) -- node [left,near start] {\; \; \;$l$ = 0}(GMGHS);
		\draw[->] (GMGHSNUT) -- node [right,near start] {$Q$ = 0}(Taub-NUT);
		\draw[->] (AKSTN) -- node [right] {\; \; \ $a$ = 0}(GMGHSNUT);
		\draw[->] (AKSTN) -- node [left] { $b$ = 0 \,\,\,}(KSTN);
		\draw[->] (AKS) -- node [right] { $a$ = $0$}(AGMGHS);
		\draw[->] (KSTN) -- node [left] {\; $l=0$}(Kerr-Sen);
		\draw[->] (Taub-NUT) -- node [right] {\; $l=0$}(Sch);
		\draw[->] (AGMGHS) -- node [ right,near start] {$b$ = 0} (GMGHS);
		\draw[->] (Kerr-Sen) -- node [right] {\; \; $Q$ = 0} (Kerr);
		\draw[->] (GMGHS) -- node [left, near start] {\; \; $Q=0$}(Sch);
		\draw[->] (Kerr) -- node [left,near start] {$a=0$ \qquad }(Sch);
		\end{tikzpicture}
		\vspace{0.1cm}
		\caption{Map of the family of accelerating and rotating spacetimes with NUT charge in the low energy limit of heterotic string theory.}
		\label{fig:map}
		
	\end{center}
\end{figure}
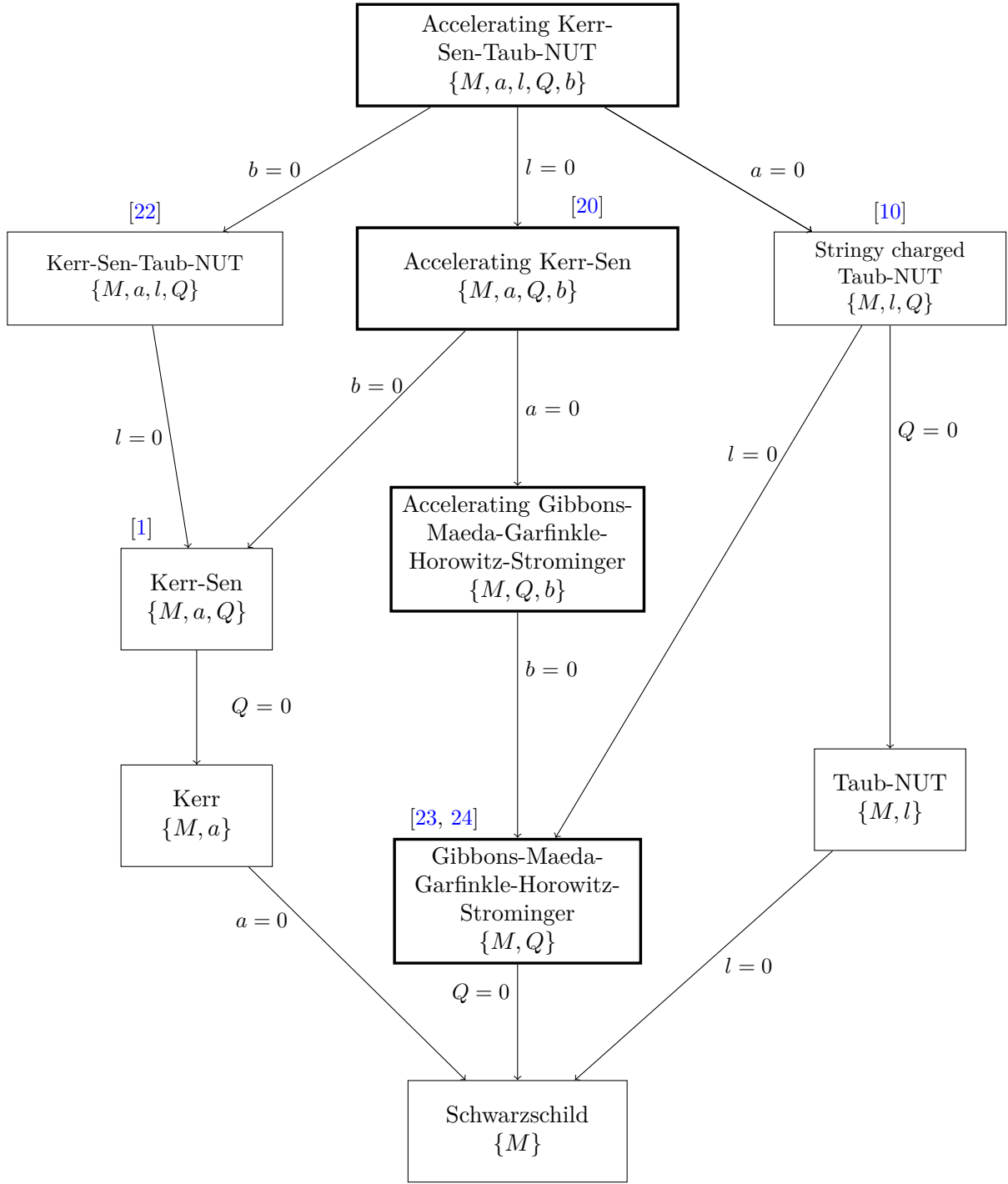

\section{Some aspects of the solution}\label{sec.aspects}

\subsection{Locations of horizons}

In the preceding section, we made reference to the black hole horizons denoted as $r_\pm$. Similar to the AKNTN spacetime, the AKSTN geometry also encompasses additional horizons arising from the acceleration parameter $b$. The locations of these horizons correspond to the roots of $\Delta_r$ in the metric (\ref{metric.AKSTN.Einstein}). The four horizons within the AKSTN spacetime can be written as the followings
\begin{eqnarray}
	{\cal H}_{bh}^+ \quad \hbox{at} \quad r_{+}
	&=& \left(M-\frac{Q^2}{2M} \right) +\sqrt{\left(M-\frac{Q^2}{2M} \right)^2 + l^2 - a^2 }\,, \label{r+rep}\\[1mm]
	{\cal H}_{bh}^- \quad \hbox{at}\quad r_{-}
	&=& \left(M-\frac{Q^2}{2M} \right) -\sqrt{\left(M-\frac{Q^2}{2M} \right)^2 + l^2 - a^2}\,, \label{r-rep}\\[2mm]
	{\cal H}_{ac}^+ \quad \hbox{at} \quad  r_a^+ &=& +\frac{1}{b}\,\frac{a^2+l^2}{a^2+a\,l}\,, \label{ra+}\\
	{\cal H}_{ac}^- \quad \hbox{at} \quad  r_a^- &=& -\frac{1}{b}\,\frac{a^2+l^2}{a^2-a\,l}\,. \label{ra-}
\end{eqnarray}
The horizons denoted as ${\cal H}_{bh}^\pm$ pertain to the black hole, while ${\cal H}_{ac}^\pm$ are associated with the acceleration. This arrangement of horizons mirrors the structure found in the AKNTN spacetime, which has been extensively elaborated upon in \cite{Podolsky:2021zwr}. Here, we will simply reiterate some properties that bear a resemblance to the discussions regarding the AKNTN geometry. 

The extreme condition is denoted by the overlapping of ${\cal H}_{bh}^+$ and ${\cal H}_{bh}^-$, i.e.
\be 
r_+ = r_- = M-\frac{Q^2}{2M}\,.
\ee 
When the hyperextreme state is reached, meaning that condition (\ref{eq.BHcondition}) is no longer met, the black hole horizons cease to exist within the spacetime. As for the acceleration horizons, both vanish when the acceleration parameter tends to zero, specifically when $r_a^\pm$ approaches positive and negative infinity as $b$ approaches zero. However, for the acceleration horizon denoted as ${\cal H}_{ac}^-$, it also disappears when $a=l$. In this particular scenario, the remaining horizons within the spacetime reduce to two: the black hole horizon at $r_+ = 2 (M-b)$ and the outer acceleration horizon at $r_a^+ = b^{-1}$.

\subsection{Ergoregions}

Ergoregion in AKSTN spacetime is denoted by $G_{tt}=0$, where explicitly we have
\be 
G_{tt} = -\frac{\left(\Delta_r - P a^2 \Delta_x\right)}{\Lambda \Omega^2 \Sigma}\,.
\ee 
Although the metric function presented above appears intricate, the condition for the vanishing of $G_{tt}$ is solely determined by
\be \label{eq.ergo}
\Delta_r - P a^2 \Delta_x = 0\,.
\ee 
Satisfying the last condition ensures $\Lambda = 1+ \sinh^2 \alpha$ to be definite positive. Note that at the poles $x= \pm 1$, the solution to the last equation is the black hole horizon radii. If we set the acceleration parameter to zero, we can find the solutions to (\ref{eq.ergo}) as
\be 
r_e  = \left( {M - \frac{{Q^2 }}{{2M}}} \right) + \sqrt {\left( {M - \frac{{Q^2 }}{{2M}}} \right)^2  + l^2  - a^2 x^2 } \,.
\ee 

However, eq. (\ref{eq.ergo}) presents a complex problem, requiring us to solve for the radius, which depends on the parameter $x$ and represents the ergosurface. Hence, to examine the alterations in the ergosurface resulting from variations in the acceleration parameter and the contributions of the NUT parameter, we provide some of numerical evaluations below. Certain patterns observed in the plots exhibit similarities to those found in the AKNTN spacetime, as discussed in \cite{Podolsky:2021zwr}. In all of the numerical examples, we use $Q=M$ and $a=0.5~\left(M-\tfrac{Q^2}{2M}\right)$. 

\begin{figure}[H]
	\begin{center}
		\includegraphics[scale=0.35]{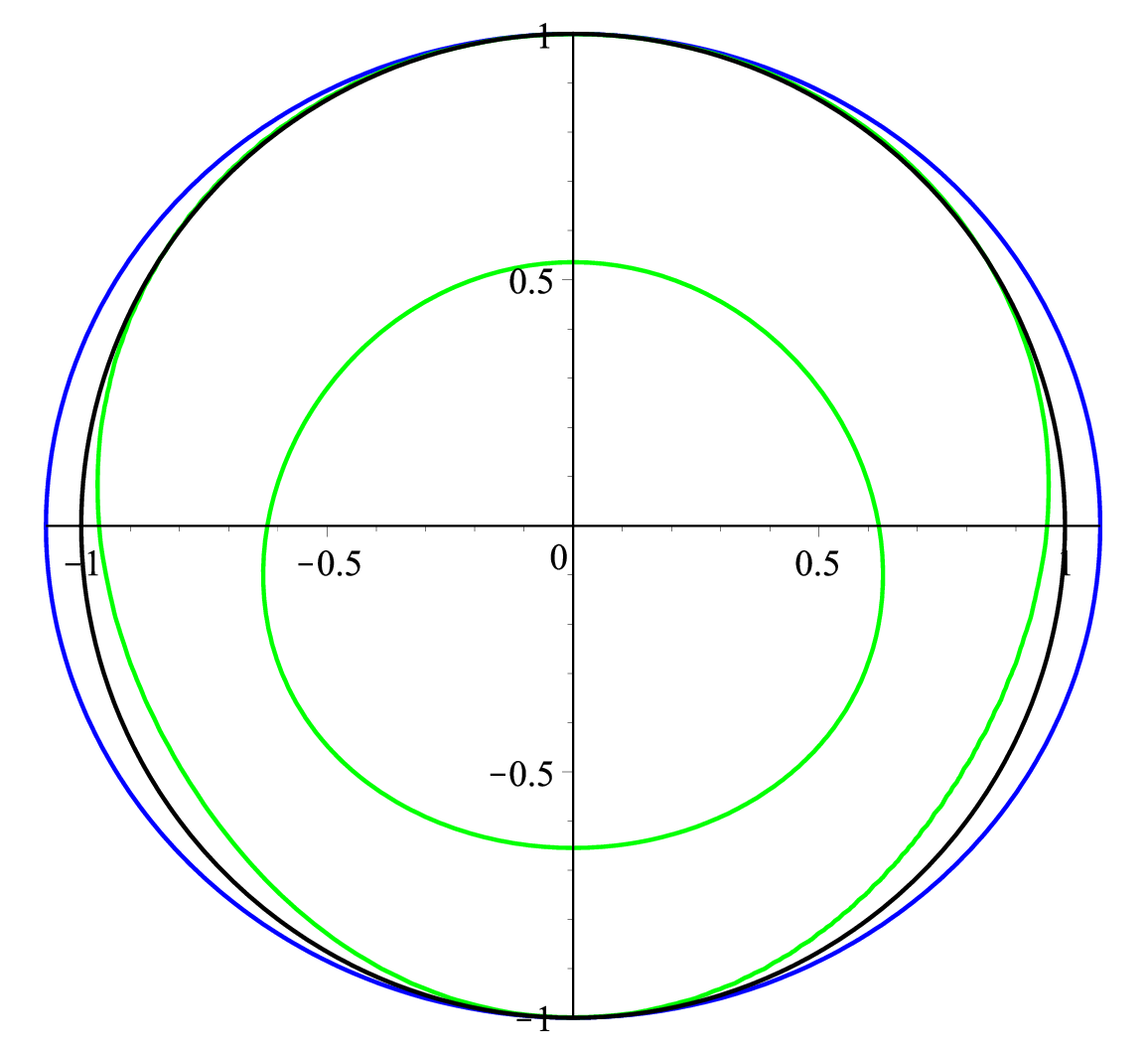}\caption{The case of null NUT charge. Black curve is the outer black hole horizon, and blue one is the associated outer ergosurface for non-accelerating case. The green curves correspond to the case with accelerating parameter $b \left(M-\tfrac{Q^2}{2M}\right)=1$. The larger one is the outer ergosurface, whereas the smaller one corresponds to the inner one.}\label{fig.ergo0}
	\end{center}
\end{figure}

\subsection{Conical singularities}

Since the initial seed solution exhibits a conical singularity, which is known to arise from the acceleration and NUT parameters, it is evident that the Hassan-Sen transformed solution discussed in this paper similarly inherits this issue. For instance, one can assess the following quantity
\be
C = \mathop {\lim }\limits_{x \to  + 1} \frac{2\pi}{{\Delta _x }}\sqrt {\frac{{G_{\phi \phi } }}{{G_{xx} }}} \,.
\ee 
If the spacetime with metric tensor $G_{\mu\nu}$ is free from the conic singularity, then we should get $C=2\pi$, i.e. one round of $\phi$ coordinate. However, for the AKSTN spacetime, either in string frame ${\tilde g}_{\mu\nu}$ or Einstein frame $G_{\mu\nu}$, one can find
$ C = 2\pi C_+$ where
\be 
C_+ = \frac{a^2(a-l)(a+l)^3 b^2 - 2amb(a+l)(a^2+l^2)+(a^2+l^2)^2}{(a^2+l^2)^2}\,.
\ee 
To cure the conical singularity at $x=1$, i.e. north pole, we can scale the $\phi$ coordinate as $\phi \to C_+^{-1}\phi$, which yields this coordinate to be regular at north pole. However, it leaves the south pole to still suffer the conical singularity. It can be seen by evaluating 
\be
\mathop {\lim }\limits_{x \to  - 1} \frac{2\pi}{{\Delta _x }}\sqrt {\frac{{G_{\phi \phi } }}{{G_{xx} }}} \,,
\ee 
which turns out to be quite lengthy, which we do not provide it here. Nevertheless, the key takeaway here is that the conical singularity present in the AKSTN spacetime cannot be straightforwardly eliminated by merely scaling $\phi$. While one can utilize a time coordinate transformation, as demonstrated in \cite{Podolsky:2021zwr}, it introduces a periodic new time coordinate, leading to a separate issue. The existence of the conical singularity in AKSTN can be understood as a reflection of the presence of a string or strut responsible for the acceleration to the black hole.

\subsection{Closed timelike curves}

\begin{figure}[H]
	\begin{center}
		\includegraphics[scale=0.5]{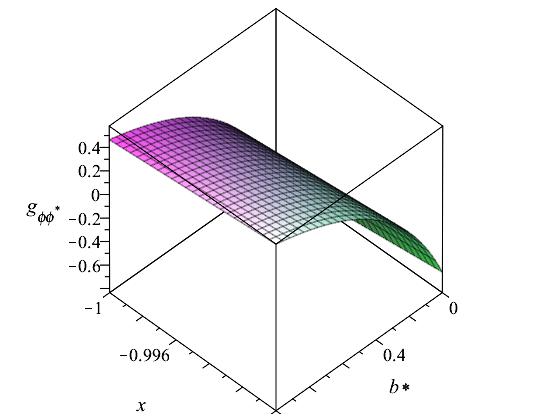}\caption{Here we use $Q=M$, $a=0.5~\left(M-\tfrac{Q^2}{2M}\right)$, and $l=0.4~\left(M-\tfrac{Q^2}{2M}\right)$. The dimensionless quantity $b^* = b \left(M-\tfrac{Q^2}{2M}\right)$.}\label{fig.ctc}
	\end{center}
\end{figure}

Once more, the presence of closed timelike curves (CTCs) in the AKSTN spacetime is another anticipated feature, as it is already present in the seed solution during the Hassan-Sen transformation. In this instance, the concept of CTC is linked to the change in the signature of $G_{\phi\phi}$ to become negative, thus rendering it timelike. Note that here we consider the mostly positive signature of spacetime. The closed nature of these curves is due to the periodicity of the $\phi$ coordinate, which does not necessarily have to be confined to a $2\pi$ period. However, given the intricacy of $G_{\phi\phi}$, we find it more appropriate to illustrate the transition from a spacelike to a timelike nature through numerical means  in Figure \ref{fig.ctc}. The figure reveals that an increase in the acceleration parameter has the capability to eliminate the closed timelike curves (CTCs) within the analyzed region.

\subsection{Area-temperature product}

The connection between the area of a horizon and its entropy is a well-established concept. The discussion of entropy here extends beyond black hole horizons alone. Several studies have indicated that entropy can also be attributed to cosmological horizons, even in the context of de Sitter space, as explored in \cite{Siahaan:2023kcz, Siahaan:2022ecb}. Consequently, we can ascribe entropy to an accelerated horizon, such as the one in the AKSTN spacetime discussed in this paper. The relationship between entropy and the horizon is as follows 
\be 
S_h = \frac{A_h}{4}\,,
\ee 
which can apply to the black hole and acceleration horizons in the spacetime. 
In this context, the subscript ``$h$'' can refer to both black hole and acceleration horizons.

In this section, we do not consider the scaling of the angular coordinate $\phi$, which is introduced to rectify the conical singularity at one pole. This is because resolving such a singularity at one pole does not eliminate it at the other. However, our aim is to confirm a relationship that exists between entropies in this spacetime, specifically the area product. The area of the horizon with a radius of $r_h$ can be calculated using the standard formula
\be 
{\cal A}_h =
\int\limits_{x =  - 1}^1 {\int\limits_0^{2\pi } {{\left. {\sqrt {G_{xx} G_{\phi \phi } } } \right|_{r=r_h } } dxd\phi } } \,.
\ee 
The result can be found as
\be \label{eq.area}
{\cal A}_h = \frac{{4\pi \left( {r_h^2  + \left( {a + l} \right)^2 } \right)\left( {1 + \sinh ^2 \alpha } \right)}}{{\left( {1 - br_h \frac{{a^2  + al}}{{a^2  + l^2 }}} \right)\left( {1 + br_h \frac{{a^2  - al}}{{a^2  + l^2 }}} \right)}}\,.
\ee 
When the NUT parameter $l$ is absent, the final outcome simplifies to the area of the horizon in the accelerating Kerr-Sen spacetime, as presented in \cite{Siahaan:2018qcw}, with the exception of certain scaling constants. However, it is important to note that the area (\ref{eq.area}) is finite for the inner and outer black hole horizons only, while it becomes singular when considering the acceleration radii $r_a^+$ and $r_a^-$.

In this section, we aim to establish a general relationship that has been observed in various instances of rotating and charged spacetimes. This relationship pertains to the product of entropy and temperature, or area and surface gravity, and it holds true for both inner and outer black hole horizons in the following form
\be \label{eq.areatempPROD}
{\cal A}_{bh}^+ T_{H}^+ = - {\cal A}_{bh}^- T_{H}^- = {\rm constant}\,.
\ee 

In the last equation, it should be noted that ${\cal A}_{bh}^+$ and ${\cal A}_{bh}^-$ represent the areas of the outer and inner black hole horizons, respectively, while $T_{H}^+$ and $T_{H}^-$ denote their corresponding Hawking temperatures. However, due to the intricate nature of the metric function in the AKSTN spacetime, calculating the Hawking temperature requires the use of surface gravity $\kappa$ in a spacetime characterized by the Killing vector $\xi^\mu$,
\be 
\kappa ^2  =  - \left( {\nabla _\mu  \xi _\nu  } \right)\left( {\nabla ^\mu  \xi ^\nu  } \right)
\ee 
is not a simple calculation. 

Nevertheless, there are several techniques available to calculate the Hawking temperature. One such approach is the tunneling method, which has been employed in numerous cases involving charged and rotating spacetimes \cite{Parikh:1999mf,Srinivasan:1998ty,Akhmedov:2006pg}. In the context of rotating spacetime, the procedure involves computing the tunneling process along the rotational axis, specifically when $x=1$, while taking into account radial and null geodesics. In this analysis, only the effective spacetime metric components $G_{tt}$ and $G_{rr}$ are relevant. As a result, the general formula for the Hawking temperature becomes
\be 
T_H^{\pm}  = \left. {\frac{{\sqrt { - \left( {\partial _r G_{tt} } \right)\left( {\partial _r G^{rr} } \right)} }}{{4\pi }}} \right|_{r = r_\pm } \,.
\ee 
Using this formula, the Hawking temperature can be read as
\be \label{eq.temp}
T_H^ \pm   = \frac{{{c_0  + c_1 b + c_2 b^2 } }}{{2\pi \left( {r_ \pm ^2  + \left( {a + l} \right)^2 } \right)\left( {a^2  + l^2 } \right)^2 \left( {1 + \sinh ^2 \alpha } \right)}}
\ee 
where
\[
c_0 = (a^2+l^2)^2(r_\pm -m)\,,
\]
\[
c_1 = al(a^2+l^2)(4r_\pm m + l^2-a^2 -3 r_\pm^2)\,,
\]
\[
c_2 = a^2 r_\pm (a^2-l^2)(3 r_\pm m + l^2-a^2-2 r_\pm^2)\,.
\]
Multiplying the Hawking temperature in eq. (\ref{eq.temp}) and the area of black hole horizon given in eq. (\ref{eq.area}), we can find
\be \label{eq.AreaTempProdAKSTN}
{\cal A}_{bh}^+ T_{H}^+ = -{\cal A}_{bh}^- T_{H}^-  = \frac{{2\left( {d_0  + d_1 \delta } \right)}}{{\left( {ba\left( {m + \delta } \right)\left( {a - l} \right) + a^2  + l^2 } \right)\left( {ba\left( {m + \delta } \right)\left( {a + l} \right) - a^2  - l^2 } \right)}}
\ee 
where
\[
d_0 = 2ba (a^2-m^2-l^2)(a^3mb+la^2-l^2bma+l^3)\,,
\]
\[
d_1 = a^2 (a^2-l^2)(a^2-2m^2-l^2)b^2-2alm (a^2+l^2)b + (a^2+l^2)\,,
\]
and $\delta = \sqrt{m^2+l^2-a^2}$. 
This outcome confirms the area-temperature product mentioned in equation (\ref{eq.areatempPROD}). It is not particularly surprising that this product relationship remains valid in the AKSTN spacetime, given that this spacetime is a Hassan-Sen transformed solution with the accelerating Kerr-Taub-NUT spacetime serving as the seed. In a previous study by Podolsky and others \cite{Podolsky:2021zwr}, it was demonstrated that this product relationship is satisfied in the seed solution. In fact, the relation ${\cal A}_{bh}^+ T{H}^+ = {\cal A}_{bh}^- T{H}^- $ is equivalent to the statement that the area product between the black hole horizons ${\cal A}_{bh}^+ {\cal A}_{bh}^-$ is independent of the black hole mass \cite{Chen:2012mh}. The result presented in equation (\ref{eq.AreaTempProdAKSTN}) indicates that the independence of the area product ${\cal A}_{bh}^+ {\cal A}_{bh}^-$ with respect to black hole mass is not a characteristic of the AKSTN spacetime. However, it's worth noting that this area product is not a universal property, as several cases in non-asymptotically flat spacetimes do not exhibit this feature \cite{Castro:2013pqa, Faraoni:2012je, Xu:2014qaa}. Clearly, the AKSTN spacetime is not asymptotically flat, so it is expected that this area product property does not apply in the context of the AKSTN spacetime.

\section{Conclusions}\label{sec.discussions}

In this paper, we have derived a novel solution within the low-energy limit of heterotic string theory, describing black holes that are both accelerating and rotating, and equipped with a NUT parameter. This solution, known as the AKSTN spacetime, extends the previous solution presented in \cite{Siahaan:2018qcw}, which described accelerating Kerr-Sen black holes. The AKSTN solution incorporates several parameters, including black hole mass, electric charge, rotation, NUT charge, and an acceleration parameter, as detailed in section \ref{sec.AccKSTN}, where we provided a map of the family of solutions. Section \ref{sec.aspects} discussed various aspects of the AKSTN spacetime, including the locations of horizons, the ergoregion, conical singularities, closed timelike curves (CTCs), and the area-temperature product. In many respects, the physical properties of the AKSTN spacetime bear a resemblance to the AKNTN solution discussed in \cite{Podolsky:2021zwr}. Both describe accelerating, rotating, charged black holes with NUT charge, albeit one originating from the low-energy limit of heterotic string theory, and the other being a solution to the Einstein-Maxwell equations.

For future research, several intriguing questions can be explored. One significant issue in the study of accelerating spacetimes, as also presented in \cite{Podolsky:2020xkf,Podolsky:2021zwr}, is the conical singularity. It is known that for spacetimes with NUT charge, one can introduce a constant parameter known as the Manko-Ruiz parameter, which can influence the location of the Misner string responsible for such singularities. It would be interesting to investigate if the accelerating Taub-NUT spacetime, equipped with the Manko-Ruiz parameter, can be generalized, and to see how it affects the existence of conical singularities in the spacetime. Another promising avenue for further study is the application of the Kerr/CFT correspondence \cite{Chen:2012mh} to the AKNTN and AKSTN spacetimes. We anticipate that the entropy associated with the horizons in these spacetimes can be holographically computed using the Cardy formula, offering valuable insights into their thermodynamic properties.

\section*{Acknowledgement}

This project is supported by LPPM-UNPAR. I thank the anonymous referee for his/her useful comments.
	

\end{document}